\documentclass[namedreferences]{kluwer}[1998/02/11]

\usepackage{graphicx}

\usepackage{makeidx}

\makeindex

\begin{document}
\begin{article}

\begin{opening}
\title{Scattering and absorption of light by periodic and
      \\ nearly periodic metallodielectric structures}

\author{V. \surname{Yannopapas}
\thanks{\email{vyannop@cc.uoa.gr}}
\thanks{Supported by the State
Scholarship Foundation (I.K.Y.),Greece.}}
\author{A. \surname{Modinos}}
\institute{Department of Physics, National Technical University of
Athens, Zografou Campus, GR-157~73, Athens, Greece}
\author{N. \surname{Stefanou}}
\institute{University of Athens, Section of Solid State Physics,
Panepistimioupolis, GR-157~84, Athens, Greece} \runningauthor{V.
Yannopapas et al.} \runningtitle{Scattering and absorption of
light by ...}

\date{2000/06/2}

\begin{abstract}
We consider the effect of different approximations to the
dielectric function of a silver sphere on the absorption of light
by two-dimensional and three-dimensional periodic and non-periodic
arrays of non-overlapping silver spheres in a host dielectric
medium. We present also some results on the band structure and the
absorption coefficient of light by photonic crystals consisting of
non-overlapping silver-coated spheres in a dielectric medium.
\end{abstract}
\end{opening}


Metallic particles, say spheres with diameters between 50~\AA \
and a few thousand \AA, distributed periodically or randomly on an
insulating substrate [two-dimensional (2D) systems], or many
layers of such particles [three-dimensional (3D) systems], have
interesting optical properties and have been studied for a long
time \cite{abeles}. In recent years it has become possible to
prepare such systems which are well defined (they are periodic
arrangements of same spheres) and have remarkable optical
properties. Depending on the size and distribution of the spheres,
they can be very good absorbers of light \cite{taleb}, or photonic
crystals operating as non-absorbing mirrors within a certain
frequency range of the electromagnetic (EM) spectrum \cite{zhang}.

For a theoretical analysis of the optical properties of such
systems, it is often necessary to go beyond the effective-medium
treatment described by the Maxwell Garnett (MG) approximation
\cite{bohren}. When the interparticle distance and (or) the size
of the particles become(s)comparable with the wavelength of the EM
radiation, or when the fractional volume occupied by the spheres
is larger than 0.3 or so, the MG theory breaks down and one needs
to solve Maxwell's equations accurately for a proper description
of the optical properties of these systems. In recent years, and
mainly in relation to the study of photonic crystals, a number of
methods have been developed which allow one to do so
\cite{soukoulis}.

Using our method of calculation \cite{modinos,skm}, we calculated
the optical properties of periodic systems: 2D arrays of metallic
spheres on a substrate \cite{metal_2} and thicker slabs consisting
of many layers of spheres \cite{optical}. We have also studied the
effect of moderate disorder on the optical properties of the above
systems, using a variation of the coherent potential approximation
(CPA) \cite{cpa2d,cpa3d}. In most of the above papers we have
assumed that the metallic spheres are plasma spheres, i.e. the
optical properties of the isolated (single) sphere are described
by a dielectric function
\begin{equation}
\epsilon_{p}(\omega)= 1-\frac{\omega_{p}^{2}}
{\omega(\omega+\mathrm{i}\tau^{-1})} \label{drude}
\end{equation}
where $\omega_{p}$ stands for the bulk plasma frequency of the
metal and $\tau$ is the relaxation time of the conduction-band
electrons. In the present paper we examine more closely the role
of the optical properties of the single sphere in the
determination of the optical properties of the composite system.

The dielectric function $\epsilon(\omega)=\epsilon_{1}(\omega)+
\mathrm{i}\ \epsilon_{2}(\omega)$ for bulk silver has been
determined experimentally by Johnson and Christy \cite{johnson}
over the range from $\hbar\omega=0.64$~eV to $\hbar\omega=6.60$
eV. With $\hbar\tau^{-1}=\hbar\tau_{b}^{-1}=0.02$~eV and
$\hbar\omega_{p}=9.2$~eV  [which is obtained from the well known
formula: $\omega_{p}^{2}=(4 \pi e^{2} n)/m$], Eq.~(\ref{drude})
reproduces satisfactorily the experimentally determined
$\epsilon_{1}(\omega)$ over the whole of the above frequency
range, and the experimentally determined $\epsilon_{2}(\omega)$ at
low frequencies ($\hbar\omega<1.5$~eV). $\tau_{b}$ refers to the
collision time of conduction-band electrons in  the bulk metal;
the collision time of an electron in a sphere of radius $S$ may be
approximated by: $\tau^{-1}=\tau_{b}^{-1}+v_{F}S^{-1}$, where
$v_{F}$ is the average velocity of an electron at the Fermi
surface. For silver $v_{F}= 1.4\cdot10^{6}$~m/sec, so that for a
sphere of radius $S=50$~\AA, using $\hbar\tau_{b}^{-1}=0.02$~eV,
we obtain $\hbar\tau^{-1}=0.2$~eV. Therefore, a more realistic
description of the dielectric function of a silver sphere would be
\cite{abeles}
\begin{equation}
\epsilon_{S}(\omega)=\epsilon(\omega)+\frac{\omega_{p}^{2}}
{\omega(\omega+\mathrm{i}\ \tau_{b}^{-1})}- \frac{\omega_{p}^{2}}
{\omega(\omega+\mathrm{i}\ \tau^{-1})} \label{ecorr}
\end{equation}
where $\epsilon(\omega)$ is the experimentally determined
dielectric function of bulk silver. We note that
$\epsilon_{S}(\omega)$ as given above, is different from the
approximation to the dielectric function of the silver sphere
obtained from Eq.~(\ref{drude}) with $\hbar\omega_{p}=9.2$~eV and
$\hbar\tau^{-1}=0.2$~eV only because the imaginary part of the
dielectric function is not represented well by Eq.~(\ref{drude})
for $\hbar\omega>1.5$~eV. We shall refer to Eq.~(\ref{drude}), as
the plasmonic approximation, to distinguish it from the more
realistic representation of Eq.~(\ref{ecorr}).
\begin{figure}
\centerline{\includegraphics[width=22pc] {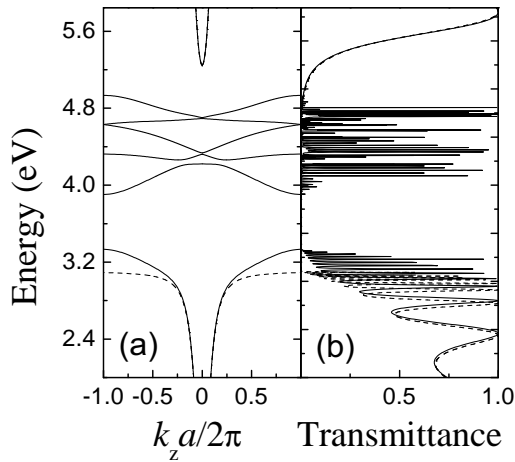}} \caption
{(a): The band structure normal to the (001) surface  of an fcc
crystal of plasma spheres ($S=50$~\AA, $\hbar\omega_{p}=9.2$~eV,
$\tau=\infty$) in gelatine ($\epsilon=2.37$), with $f=0.3$. (b):
The corresponding transmittance curve for light incident normally
on a slab, 16-layers thick, of the above crystal. The solid
(broken) lines in both figures represent the exact
(effective-medium) results.} \label{fig1}
\end{figure}

We consider a crystal of identical silver spheres centered on the
sites of an fcc lattice, with lattice constant $a$. We view the
crystal as a stack of layers (planes of spheres) parallel to the
(001) surface (which we assume parallel to the $xy$ plane). For
any given ${\bf k}_{\parallel}=(k_{x},k_{y})$, we can calculate
the real frequency lines: $k_{z}=k_{z}(\omega,{\bf
k}_{\parallel})$. For any given ${\bf k}_{\parallel}$, there may
be, for a few frequency lines, regions of $\omega$ over which
$k_{z}$ is real. These regions define the frequency bands of the
EM field in the infinite crystal. In the absence of absorption
($\tau=\infty$)  the band structure of a photonic crystal
consisting of silver spheres in a non-absorbing host medium is
determined solely by the real part of the dielectric function and
is, therefore, the same whether we use Eq.~(\ref{drude}) or
Eq.~(\ref{ecorr}) to describe the spheres. In Fig.~\ref{fig1}a we
show the band structure, for ${\bf k}_{\parallel}={\bf 0}$ [normal
to the (001) surface], for a crystal where the fractional volume
occupied by the spheres $f=0.3$. The broken lines show the bands
that are obtained using the MG approximation. The flat bands,
which are not obtained in the effective-medium treatment, derive
from $2^{l}$-pole resonances ($l>1$) of the individual spheres.
Next to the band structure, in Fig.~\ref{fig1}b, we show the
transmission coefficient of light incident normally on a slab of
the crystal consisting of 16 planes of spheres parallel to the
(001) surface. The medium on either side of the slab has the same
dielectric constant $\epsilon=2.37$ as between the spheres of the
slab. We see that the exact results agree with those of the MG
treatment except in the region of the multipolar bands which, in
the present case, are prominent as they fall within the frequency
gap of the MG band structure. A detailed discussion of the band
structure shown in Fig.~\ref{fig1} can be found elsewhere
\cite{optical}.

\begin{figure}
\centerline{\includegraphics[width=17pc] {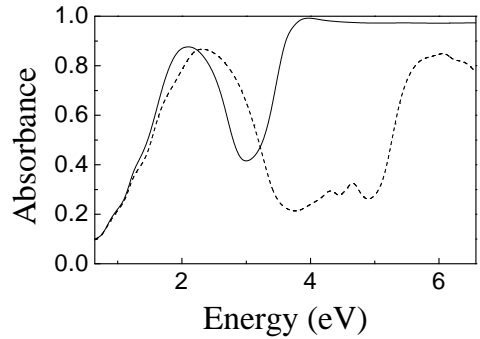}}
\caption{Absorbance of light incident normally on a 129-layers
thick slab of the crystal  described in the caption of
Fig.~\ref{fig1}, except that $\hbar\tau^{-1}=0.2$~eV. The solid
(broken) line is obtained with the realistic (plasmonic)
dielectric function of the silver spheres.} \label{fig2}
\end{figure}

We now turn our attention to the absorption coefficient of light
which involves the imaginary part of the dielectric function of
the silver spheres which, as we have seen, differs from that of
plasma spheres. In Fig.~\ref{fig2} we show the absorption
coefficient of light incident normally on a slab of the  above
crystal, 129-layers thick. The dip in the absorbance, in the
region from $\hbar\omega\simeq3.2$~eV to
$\hbar\omega\simeq5.2$~eV, corresponds to the frequency gap of the
dipolar bands shown in Fig.~\ref{fig1}a, and two relatively small
peaks at about $\hbar\omega\simeq4.5$~eV are obviously the result
of weak absorption by the multipolar bands, at about the same
frequency, shown in Fig.~\ref{fig1}a. When the more realistic
approximation of Eq.~(\ref{ecorr}) is used, the dip shrinks and
with it disappears the fine structure associated with the
multipolar bands, as shown by the solid line of Fig.~\ref{fig2}.
We note that the shrinking is effected asymmetrically and that the
minimum of absorption now occurs at a lower frequency. We may add
that the result represented by the solid line of Fig.~\ref{fig2}
does not differ significantly from that obtained from the MG
theory, when we use in the latter the dielectric function of
Eq.~(\ref{ecorr}). This, however, is not generally true.

\begin{figure}
\centerline{\includegraphics[width=22pc] {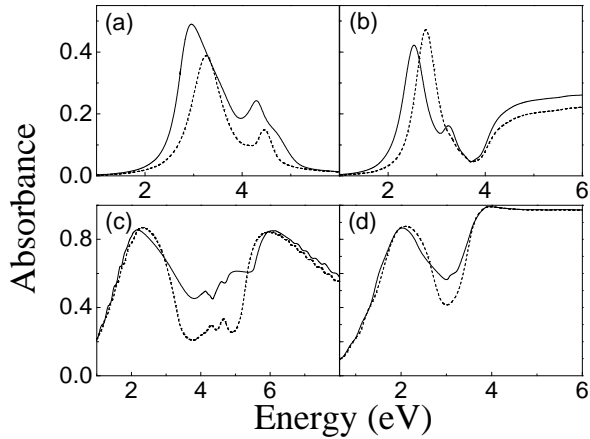}}
\caption{Absorbance of light incident normally on a (001) slab of
an fcc crystal of silver spheres ($S=50$~\AA,
$\hbar\omega_{p}=9.2$~eV, $\hbar\tau^{-1}=$0.2~eV) in gelatine
($\epsilon=2.37$). Slab of one layer: (a) plasmonic, (b) realistic
dielectric function. Slab of 129 layers: (c) plasmonic, (d)
realistic dielectric function. Solid lines: random occupancy of
75\% of the lattice sites. Broken lines: ordered system with the
same volume (surface) coverage by the spheres [30\% (43\%)] as for
the disordered 3D(2D) system.} \label{fig3}
\end{figure}

In Fig.~\ref{fig3} we compare the absorbance by different systems
as calculated using the plasmonic approximation for the single
sphere with more realistic results based on Eq.~(\ref{ecorr}). The
two top diagrams, (a) and (b), refer to a single plane of spheres.
The broken lines correspond to an ordered arrangement: the spheres
are centred on a square lattice with a lattice constant
$a_{0}=135.14$~\AA. The solid lines correspond to a disordered
arrangement: the spheres occupy at random a fraction $c=0.75$ of
the sites of a square lattice with a lattice constant
$a'_{0}=a_{0}\sqrt{c}$, so that the average number of spheres per
unit area is the same as for the ordered arrangement. The
calculation of the absorbance for the disordered arrangement was
done using a CPA formalism \cite{cpa2d}.
In relation to the ordered arrangements (2D and 3D) of plasma
spheres we observe that the lower-frequency peak in the
absorbance, which is due to dipolar absorption by the plasma
oscillations is shifted slightly toward higher frequencies and it
is wider than that obtained with the more realistic approximation
of Eq.~(\ref{ecorr}). Fine structure in the absorbance, which
occurs at about $\hbar\omega\simeq4.5$~eV in the plasmonic
approximation, and which is due to multipolar absorption by the
spheres, disappears altogether with the more realistic dielectric
function of Eq.~(\ref{ecorr}). This appears to explain why this
fine structure has not been detected experimentally \cite{taleb}.
Finally, the absorption coefficient increases to an almost
constant value at higher frequencies, $\hbar\omega>4$~eV, when the
spheres are described by Eq.~(\ref{ecorr}). In contrast the
absorption coefficient diminishes to almost zero for
$\hbar\omega>4$~eV in the plasmonic approximation.

In relation to the disordered 2D system, represented by the solid
lines in Figs.~\ref{fig3}a and \ref{fig3}b, we note the following.
In the plasmonic approximation, disorder leads to increased
absorption at all frequencies. When the spheres are described by
Eq.~(\ref{ecorr}) disorder leads to the appearance of an
additional peak at $\hbar\omega\simeq3.2$~eV where none existed
for the ordered arrangement. The dipolar peak shifts to lower
frequencies by about 0.2~eV, but the absorption associated with it
is reduced in contrast with what happens in the plasmonic
approximation. At higher frequencies, $\hbar\omega>4$~eV, disorder
leads to some further increase in absorption.

Figs.~\ref{fig3}c and \ref{fig3}d refer to a (001) fcc slab,
129-layers thick. The ordered slab is viewed as a succession of
the periodic planes of spheres considered in Figs.~\ref{fig3}a and
\ref{fig3}b, which corresponds to $f=0.3$. In the disordered
structure the spheres occupy 75\% of the sites of an fcc lattice
with a lattice constant $a=173.65$~\AA\ so that there is on
average the same volume coverage by the spheres as in the ordered
structure. The absorbance of the disordered 3D structure was
calculated using a CPA method \cite{cpa3d}. The physics underlying
the absorption spectra of the 3D systems has been discussed in
detail elsewhere \cite{optical} in relation to the ordered
structure, using the plasmonic approximation, represented by the
broken line in Fig.~\ref{fig3}c. The main peaks in the said
absorbance curve, at $\hbar\omega\simeq2.28$~eV and
$\hbar\omega\simeq5.96$~eV, are dipolar in origin and are
associated with the plasma resonances of the individual spheres.
The interaction between the spheres and with the host medium opens
up the frequency gap shown in Fig.~\ref{fig1}, and dipolar
absorption, therefore, occurs mainly at and about the edges of
this gap, leading to the two peaks mentioned above. The two lesser
peaks, in between the above two, exhibited by the broken line of
Fig.~\ref{fig3}c are due to multipolar absorption by corresponding
bands shown in Fig.~\ref{fig1}a. The solid line in
Fig.~\ref{fig3}c shows that, in the plasmonic approximation,
disorder increases the absorbance in the region between the
dipolar peaks, but it does not remove the fine structure due to
multipolar peaks. If anything, it adds to it; the additional
structure derives from the fact that one deals with effective
non-spherical scatterers in the CPA treatment of disorder
\cite{cpa3d}. In Fig.~\ref{fig3}d we show the absorbance by a slab
of the material, the same as that of Fig.~\ref{fig3}c, except that
the spheres are now described by Eq.~(\ref{ecorr}). In this case
the dip in dipolar absorption, associated with the frequency gap
of Fig.~\ref{fig1}, is compressed to lower frequencies, the fine
structure in the dip practically disappears and the absorption
coefficient increases to near unity for $\hbar\omega>4$~eV.

We note that introducing disorder in a 2D system leads to a
considerable shift and broadening of the main absorbance peaks,
but in a 3D system the effect of disorder seems to be marginal
with the exception of an increased absorbance in the dip between
the dipole peaks.
\begin{figure}
\centerline{\includegraphics[height=10.7pc] {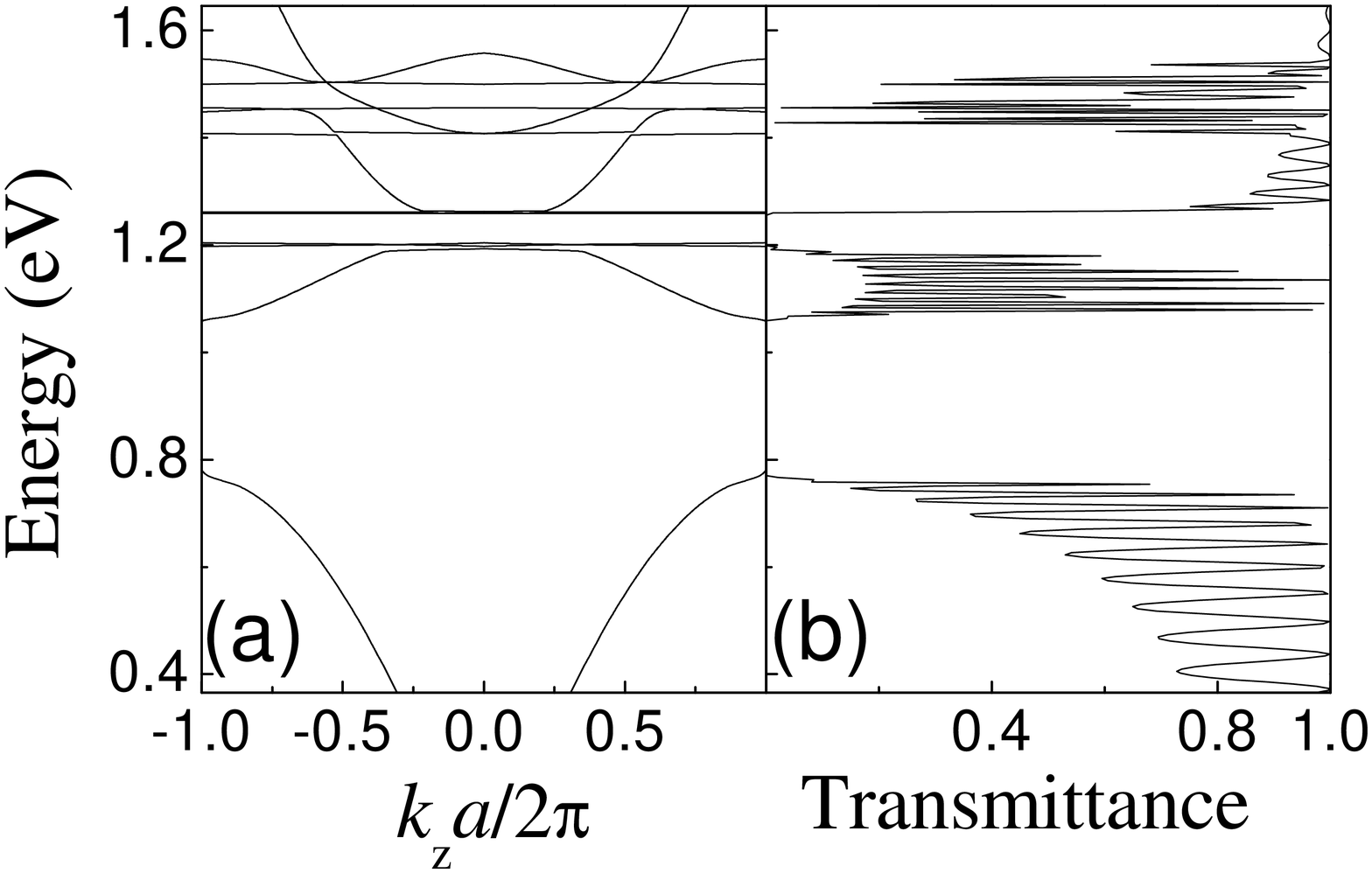}
\includegraphics[height=10.7pc] {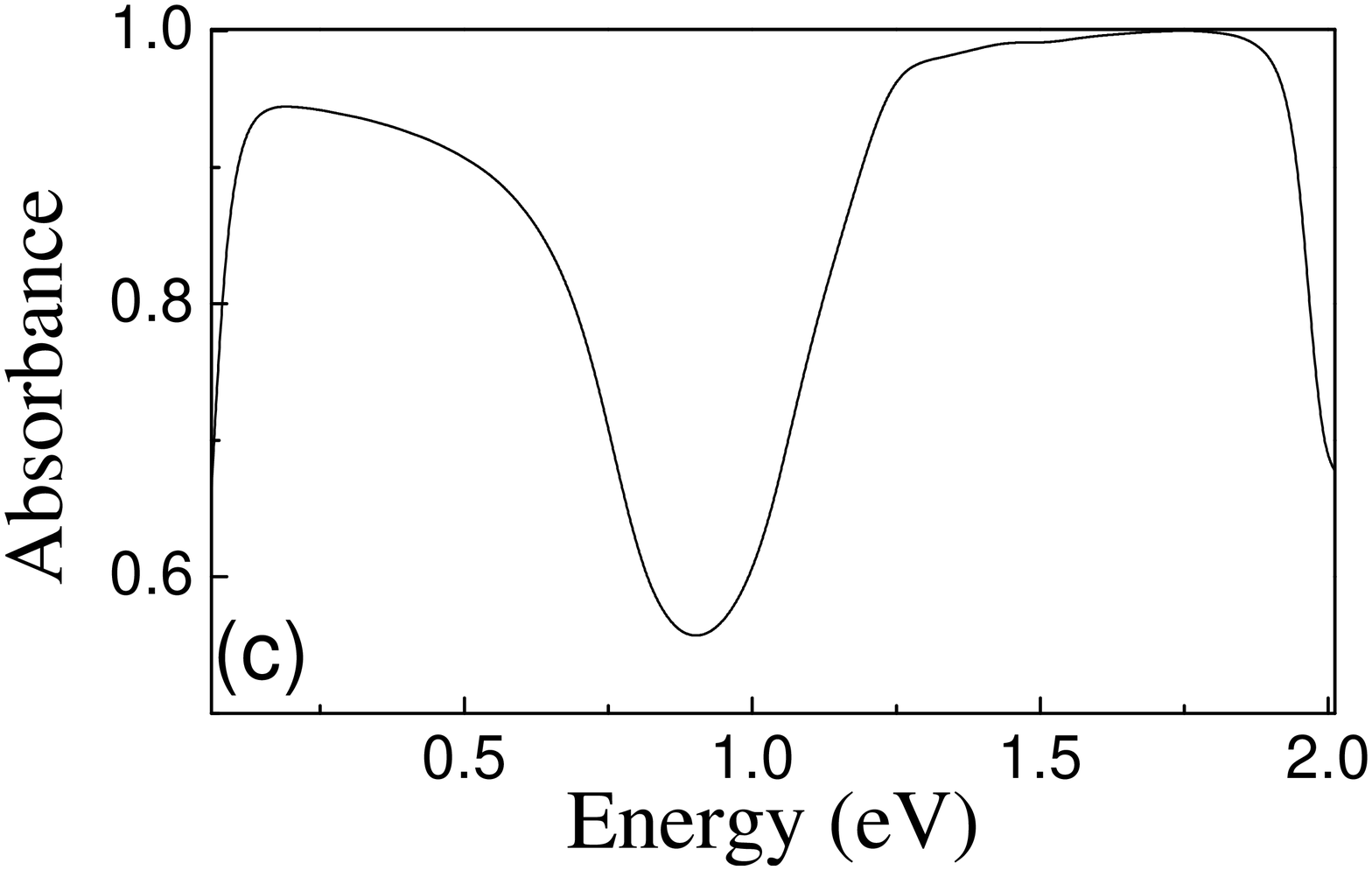}} \caption {(a): The photonic
band structure normal to the (001) surface  of an fcc crystal of
silver-coated spheres ($S_{c}=1054.8$~\AA, $\epsilon_{c}=2.37$,
$D=50$~\AA, $\hbar\omega_{p}=9.2$~eV, $\tau=\infty$) in gelatine
($\epsilon=2.37$), with $f=0.3$. (b): The corresponding
transmittance curve for light incident normally on a (001) slab,
16-layers thick, of the above crystal. (c): Absorbance of light
incident normally on a 129-layers thick slab of the above crystal,
except that $\hbar\tau^{-1}=0.2$~eV.} \label{fig4}
\end{figure}

Finally, we consider a 3D array of non-overlapping coated spheres
centred on the sites of an fcc lattice. The radius of a coated
sphere is denoted by $S$; its core of radius $S_{c}$ consists of
dielectric material ($\epsilon_{c}=2.37$) and its outer shell of
thickness $D=S-S_{c}$ is assumed to be silver. There are two
(dipolar) plasma resonances of an isolated coated sphere in a
surrounding medium with a dielectric constant $\epsilon$ given by
\cite{bohren}
\begin{equation}
\omega_{pcs}=\frac{\omega_{p}}
{\bigl(1+K/2\mp\frac{1}{2}(K^{2}-4\epsilon\epsilon_{c})^{1/2}
\bigr)^{1/2}} \label{coatres}
\end{equation}
where $K=[\epsilon_{c}(1/2+\beta)+\epsilon(2+\beta)]/(1-\beta)$
and $\beta=(S_{c}/S)^{3}$. The frequencies $\omega_{pcs}$, given
by Eq.~(\ref{coatres}), are to be compared with the (dipolar)
plasma resonance of a homogeneous silver sphere in the same host
medium, $\omega_{ps}=\omega_{p}/\sqrt{1+2\epsilon}$. We expect
dipolar absorption by the crystal of coated spheres to occur in
the region of $\omega_{pcs}$ in the same way that absorption by a
crystal of silver spheres occurs in the region of $\omega_{ps}$.
We note that the value of $\omega_{pcs}$ corresponding to the plus
sign of Eq.~(\ref{coatres}), when $\beta$ is close to unity, is
much smaller than $\omega_{ps}$, and, as we have already noted, at
these frequencies ($\hbar\omega<1.5$~eV), the plasmonic
approximation is very good. The value of $\omega_{pcs}$
corresponding to the minus sign of Eq.~(\ref{coatres}) is close to
the plasma frequency of bulk silver $\omega_{p}$ and we shall not
deal with it, as it lies well above the optical region . The
scattering properties of the individual sphere enters the
calculation through the $T$-matrix for the sphere. Using the
$T$-matrix for the coated sphere \cite{bohren} in our formalism we
calculated the band structure , shown in Fig.~\ref{fig4}a, of an
fcc crystal of non-absorbing silver-coated spheres with $f=0.3$.
This is to be compared with that of Fig.~\ref{fig1}a for the
crystal of homogeneous spheres. As expected, we find that the
frequency gap associated with the dipolar plasma resonances of the
spheres opens up at lower frequencies, in the vicinity of
$\hbar\omega_{pcs}\simeq0.99$~eV. The flat bands in
Fig.~\ref{fig4}a are multipolar bands as in Fig.~\ref{fig1}a. Next
to the frequency band structure, in Fig.~\ref{fig4}b, we show the
transmission coefficient of light incident normally on a slab of
16 layers. In Fig.~\ref{fig4}c we show the absorption coefficient
of light incident  normally on a thicker slab of the same crystal,
allowing for absorption by putting $\hbar\tau^{-1}=0.2$~eV in
Eq.~(\ref{drude}). We see that the crystal of silver-coated
spheres absorbs over a frequency range much lower than that of the
crystal of homogeneous silver spheres. Which shows that by an
appropriate choice of parameters we can vary the range of
frequencies over which absorption of light takes place.

\end{article}
\end{document}